# Electronic structure of single-wall silicon nanotubes and silicon nanorribons: Helical symmetry treatment and effect of dimensionality

Pavol Baňacký[1*], Jozef Noga[2,3], Vojtech Szöcs[1]

[1]Chemical Physics Division, Institute of Chemistry and [2]Department of Inorganic Chemistry, Faculty of Natural Sciences, Comenius University, Mlynska dolina CH2, 84215 Bratislava, Slovakia
[3]Institute of Inorganic Chemistry, Slovak Academy of Sciences, Dubravska cesta 9, 84536 Bratislava, Slovakia

Helical method of tube formation for band structure calculations and Hartree-Fock self-consistent field method (HF-SCF) modified for periodic solids have been applied in study of electronic properties of single-wall silicon nanotubes (SWSiNT), graphene-like parent 2D-hP silicone sheet and nanoribbons (SiNR). The results obtained for nanotubes of the length of ≈ 358 Å in diameter range ≈ 3.7 Å – 116 Å of different helicity-types have shown that only small-diameter SWSiNTs up to $\phi$ < 6.3 Å are metallic due to the effect of curvature which induces coupling of $\sigma$ and $\pi$ orbitals. From the calculated band structures follow that irrespective of helicity, the SWSiNTs of larger diameter are all small-gap semiconductors with direct gap between the Dirac-like cones of ($\pi^*$, $\pi$) bands. Gap of SWSiNTs exhibits, however, an oscillatory-decreasing character with increase of the tube diameter. In the oscillatory series, minima of the gap in "saw-teeth" pattern are reached for helicity numbers $m_a$ that are an integer multiple of 3, whilst $m_a$ value itself directly determine the fold-number of particular tubular rotational axis symmetry. Oscillations are damped and gap decreases toward ≈ 0.33 eV for tube diameter ≈ 116 Å. Irrespective of the width, the SiNRs are all small-gap semiconductors, characteristic by oscillatory decreasing gap with increasing ribbon widths. The gap of SWSiNTs and SiNRs is tuneable through modulation of tube diameter or ribbon width, respectively. The SiNRs and SWSiNTs could be fully compatible with contemporary silicon-based microelectronics and could serve as natural junction and active elements in field of nono-micro technologies.

**PACS** number(s): 73.22.-f, 71.20.-b, 71.20.Mg, 73.22.Dj, 71.20.Gj

## I. Introduction

Discovery of 1D and 2D nanostructural form of carbon, i.e. carbon nanotubes[1] and graphene[2] with extraordinary physical properties, initiated opening of a new and rapidly growing field of research in solid-state physics and solid-state chemistry with potential applications in diverse area of nanotechnology including biological and medicinal applications. Similar electronic properties have been naturally expected for 1D and 2D nanostructure form of some other elements of group IV. In particular, in case of silicon it should be extremely important since highest possible compatibility for micro/nano junctions formation with contemporary "bulk" silicon-based microelectronic can be expected.

It is well known that sp$^2$ hybridization with strong in-plane overlap of $\pi$ orbitals is responsible for stability of 2D-hP nanostructural form of graphene. On a bulk scale it gives rise to graphite formation which is most stable crystal structure of carbon. Since interlayer interactions of $\pi$ orbitals are much weaker, it enables at certain circumstances to exfoliate even a single-layer planar sheet of carbon atoms with 2D-honeycomb pattern

---

[*] Corresponding author, E-mail: banacky@fns.uniba.sk

– reported method of graphene discovery. Carbon nanotubes formation and theirs stability is directly related to stability of graphene.

With valence (s, $p_{x,y,z}$) electrons in $3^{rd}$-shell, silicon, though nearest-neighbour of carbon in group IV, exhibits different properties. Most stable crystal form of silicon is diamond-like cF8 structure with $sp^3$ hybridization, whilst bulk form of graphite-like silicon structure is unknown. That was the main reason of uncertainty about possibility of existence and stability of $sp^2$ silicone analogue of graphene, i.e. 2D-hP single-layer Si sheet with honeycomb pattern. For a long time, an effort to prepare Si-nanostructures has resulted usually in Si nanowires[3-9] ($sp^3$-based structures), rather than to any other form. Since 2002 the first reports on synthesis of large-diameter silicon nanotubes (up to 50nm) have appeared.[10-13] Experimental evidence of tin-wall and small-diameter silicon nanotubes ($\phi \approx$ 2nm) formation[14,15] came in 2005 and the same authors reported later[16] that parts of less-oxidized Si-nanotubes possess hexagonal character which can be interpreted as a mixture of $sp^2/sp^3$ hybridization. Graphene-like pattering has also been reported[17,18] at experimental study of silicon nanoribbons (SiNR). The results of scanning tunnelling microscopy (STM) and angle-resolved photoemission spectroscopy (ARPES), published only very recently[19] have shown in a convincing way that graphene-like silicon sheet was synthesized and the authors named it silicene. More over, it has been shown[19] that single-layer sheet is lightly buckled 2D-hP structure with an average Si-Si distance $\approx$ 0.22 nm (±0.001nm) and electronic dispersion derived by ARPES confirmed presence of relativistic Dirac fermions, which is the very basic characteristic feature of graphene-like structure.

The buckled structure as a stable form of silicone (or more precisely, meta-stable allotrope of Si) has been predicted theoreticaly[20] by DFT-based study already in 1994. In combination with finite temperature molecular mechanics and DFT-based calculations it has been shown[21] that strictly planar and light buckled silicene sheet have nearly identical energy minima on adiabatic potential energy surface. For planar structure, however, there is a mixing of acoustic and optical phonon modes with lowering into acoustical region with small but imaginary frequency at Γ point whilst, in light-buckled structure are acoustic and optical branches well separated and structure is calculated to be stable up to 1000 K. For both structures, characteristic is an optical phonon mode with frequency $\approx$ 600 cm$^{-1}$ at Γ point and nearly identical topology of electronic band structures with Dirac cone at K point and Fermi velocity $\approx 10^6$ m/s. Calculated hP-lattice constant is 3.83 Å and Si-Si distance 2.25 Å. Whilst results of applied theoretical calculation methods, no matter if for strictly planar or light-buckled 2D-hP silicene sheet yields basically the same electronic dispersion, the theoretical predictions of electronic band structure for single-wall silicone nanotubes (SWSiNT) are different.

Simple metallicity condition derived by tight-binding (TB) approximation for 2D-hP planar graphene sheet and directly applied for carbon nanotubes[22], can be expressed in a simple way; for tube with chirality numbers (n,m) holds [(n-m)=3μ] and tube is metallic if (n-m) is an integer multiple of 3 and semiconducting otherwise. It has been calculated by the DFT-LDA method[23] and TB-Hamiltonian approach[24] that this relation is valid also for SWSiNTs. However, for H-terminated SWSiNTs the TB-Hamiltonian approach yields semiconductor character (gap $\approx$ 2.2 eV) for all tubes independent on chirality and diameter.[25] Semiconductor character independent on chirality but with decreasing gap with increasing tube diameter has been reported[26] for anion (silicide-like) form of

SWSiNTs and also for H-teminated SWSiNTs calculated by DFT-TB approach. Simple metallicity condition has been found to be valid[27] for SWSiNTs with strictly planar ($sp^2$) parent layer also within the DFT calculation with plane-wave basis set. The same method yields[27], however, different results for SWSiNTs with buckled ($sp^3$) character of parent layer, thought this type of tube is only of 0.03 eV more stable than corresponding $sp^2$-based tube. For studied set, irrespective of chirality, the (n,0) tubes are metallic for n=5-9 and semiconductor with decreasing gap with increasing tube diameter for n=10-24. Metallicity is ascribed to $\sigma^*$ - $\pi^*$ mixing in small diameter tubes. However, the armchair (n,n)-type tubes for n=5-11 are found to be semiconductors with decreasing gap as tube-diameter increases. The finite temperature molecular mechanics (MM) in combination with DFT-based calculations with plane-wave basis set applied in study of SWSiNTs[28] predicted the nanotubes with light-buckled parent Si 2D-hP sheet (average Si-Si distance ≈2.2 Å) to be unstable for tube diameter smaller than ≈ 7.6 Å (n<6 for (n,0) and (n,n)-types) but structure can be stabilized by internal or external adsorption of transition metal elements. The SWSiNTs (n,0)-type, irrespective of chirality, are in the range of 6≤n≥11 metallic and band gap between valence and conduction band opens for n≥12 (diameter ≈ 14.6 Å). The authors[28] suppose, however, that transition from metal to semiconductor may occur at smaller diameter if GW method of self-energy calculation is applied and, in general, DFT results may differ depending on pseudopotentials and approximation of exchange-correlation potential applied. Instability of small diameter SWSiNTs (n<6) with $sp^3$ distortion has also been calculated[29] by non-orthogonal DFT-TB in combination with MM simulation but, in contrast to the results[28], all types (zig-zag, armchair, or chiral) of SWSiNTs were found, however, to be semiconductors with small band gaps (<1 eV). Stabilization of SWSiNT by insertion of different metal atoms inside tube has been studied also by others[30] and metal character of all small-diameter SWSiNTs, regardless of chirality, has been predicted[31] also by DFT method with B3LYP/6-31G exchange-correlation potential. Clusters of different character and size have also been considered[32-34] at modelling SWSiNTs. Stability of Si-nanotubular structures has been studied by generalized TB-MM method[32] and by semiempirical HF-SCF MNDO method with PM3 parametrization.[33] The MNDO method predicted[33] that SWSiNTs with buckled ($sp^3$) structure could be a stable structure. Based on different cluster structures, aspects of electronic structure of SWSiNTs within ab-initio MP2/6-31G method have also been investigated[34] and conclusion of authors is that SWSiNTs are possibly metals rather than wide-gap semiconductors.

Experimental methods and techniques for SWSiNTs synthesis are in an early period of development. To our knowledge, so far published results of SWSiNTs synthesis reported production of mixture of different products with only a small fraction (up to 10%) of tubular structures with a wide range of diameters. Experimental characterization of electronic properties of SWSiNTs has not been published yet. In these circumstances, theoretical calculations are the only qualified source available. As presented above, calculated electronic properties of SWSiNTs are not uniform and mutually consistent. In our opinion, the reason of differences in calculated properties is simply due to the character and parametrization of applied theoretical methods, which are more or less convenient for study of particular properties, structural type and composition. It is not the goal of the present study to analyse which method is most convenient for SWSiNTs calculations. What is crucial in our opinion, however, is the *basic character of tubular*

*electronic band structure calculation*, which is the same for all band structures (BS) of SWSiNTs introduced above. The calculations are based on well-known "chiral vector" treatment of carbon nanotube formation[35,36], which is straightforwardly applied for subsequent BS calculation of tubular structure. The chiral numbers (n,m) define two orthogonal unit vectors, chiral-***C*** and translation-***T*** vector in graphene sheet and in corresponding reciprocal ***k***-space with allowed discrete values (μ) in ***C*** * direction and continuous translation $k_{tr}$-values in ***T**** direction. The rectangle defined by unit vectors ***C*** and ***T*** when rolled-up creates translation tubular unit cell for a tube with circumference |***C***| and translation modulus |***T***|. Translation tubular unit cell contains N irreducible unit cells of planar 2D-hP graphene sheet. For zig-zag (n,0) and armchair (n,n) nanotubes N=2n and number of atoms for graphene-like structures with 2 atoms in 2D-hP unit cell is 4n. In case of e.g. SWSiNT with diameter ≈ 26 Å, i.e. nanotube of (21,0)-type, the BS calculated directly for translation tubular unit cell in valence electron-basis set (4 AO/atom) is represented by a bundle of 336 bands in contrast to the BS of 2D-hP silicene sheet with only 8 bands. In case of graphene and carbon nanotubes, due to decoupling of σ and π electrons ($sp^2$ hybridization of grapheme planar 2D-hP structure), usually π-electron approximation is applied and BS of tubular structure is calculated by zone-folding method. In this case, tubular BS is derived from electronic dispersion of planar graphene $E^{2D}(k_x,k_y)$, but calculated for *k*-values (μ, $k_{tr}$) of tubular *k*-space, i.e. $E^{tb}(k) = E^{2D}(μ, k_{tr})$. This method is not restricted only for π-bands dispersion, but it is used in the same way also for complete (σ,π)-bands dispersion calculations. It is evident that zone-folding method completely neglects the effect of tube curvature, which can be substantial mainly for small-diameter nanotubes. Tubular BS calculated directly over tubular translation unit cell covers the effect of curvature in principle but, it is an open question if a method, e.g. based on a plane-wave basis set, is suitable for tubular structures calculation.

Nonetheless, no matter of theoretical method applied and approximation used, the key point in tubular BS calculation is the fact that translation tubular unit cell defined by "chiral vector" treatment is not irreducible unit cell of tubular structure. As a consequence, the correspondence between the 8n-bands BS of tubular structure defined by translation tubular unit cell and the 8-bands BS of planar parent structure defined by 2D-hP irreducible unit cell is lost and can hardly be reconstructed. More over, it gives rise to uncertainty about the basic character of calculated BS. It can be very important mainly for more complex structures with more than 2 atoms in parent 2D-hP irreducible unit cell, e.g. in case of single wall boron nanotubes (SWBNT) with 8 atoms/2D-hP u.c. It has been shown[37] that BS of SWBNT with the same chiral vector calculated by zone folding method is substantially different from the BS obtained by direct calculation based on translation tubular unit cell. Whilst the first one is metallic - particular bands intersect Fermi level, the direct calculation results in semiconductor. Similar situation can not be excluded in a graphene-like system if coupling of σ-π orbitals is induced.

In the present study, instead of commonly applied "chiral vector" treatment, we have used the helical method[38-41] of tube formation. Employing of the screw symmetry operations, helical method gives rise to preserving direct relation between BS of 2D-hP parent structure and tubular BS. Within this method, the irreducible 2D-hP unit cell of parent planar structure remains the unit cell also for tubular 1D-structure. Related to a two-dimensional structure characterized by translational vectors **a** and **b**, any helical tube can be created[42] by rolling up a ribbon corresponding to $m_a$ translations of the reference unit cell along $a$ and an "infinite" number of translations ($m_{tr}$) along $b$. The helix is defined by helical parameters ($m_a,m_b$) defining a vector ($m_a·a+m_b·b$) that is rolled up perpendicular to the helical axis. This vector is mapped on a cylinder surface, makes its circumference, and hence determines the diameter of the tube. For $[(m_a,m_b),m_{tr}]$ the reciprocal space is characterized by a pair of ($k_{\Phi r}$, $k_{tr}$)-values, $k_{\Phi r}=r/m_a$ ($r=0,1,…,m_a-1$) and $k_{tr}\in\langle-1/2,1/2\rangle$. Now, translations along $a$ correspond to rotations by $\Phi_r = 2\pi(r/m_a)$ to which $k_{\Phi r}$ is related. Translations along $b$ ($m_{tr}$) are mapped as rototranslations and can be treated as true translations in an infinite one-dimensional system, hence giving rise to continuous values for $k_{tr}$ related to reciprocal rototranslations. As a result, the BS of tubular structure is characterized by the same number of bands as the BS of parent 2D-hP structure. To model the tube in practical calculations, the rototranslations were terminated after sufficiently large odd number $m_{tr}$, which is directly related to the length of created tube ($m_{tr}>>m_b$), and a "cyclic cluster" with periodic boundary conditions was constructed at calculation of matrix elements[43] that essentially corresponds to the bulk limit.

The results presented in this paper are for nanotubes of the length of ≈ 358 Å in the diameter range ≈ 3.7 Å – 116 Å. It has been shown that of the true metallic character are only small-diameter SWSiNTs up to $\phi$ < 6.3 Å due to the effect of curvature which induces coupling of σ and π orbitals. From the calculated band structures follow that irrespective of helicity, the SWSiNTs of larger diameter are all small-gap semiconductors with direct gap between the Dirac-like cones of ($\pi^*$, π) bands. Gap of SWSiNTs exhibits, however, an oscillatory-decreasing character with increase of the tube diameter. In the oscillatory series, minima of the gap in "saw-teeth" pattern are reached for helicity numbers $m_a$ that are an integer multiple of 3, whilst $m_a$ value itself directly determine the fold-number of particular tubular rotational axis symmetry. Oscillations are damped and gap decreases toward ≈ 0.33 eV for tube diameter ≈ 116 Å. For particular tubular structures, besides the band structure and corresponding gap, excitation energy and energy of folding (stretch energy) is also calculated. The results for SiNRs show that irrespective of the width, all studied ribbons are small-gap semiconductors characteristic by oscillatory decreasing gap with increasing ribbon widths. From the results follows that gap of SWSiNTs and SiNRs is tuneable through modulation of tube diameter or ribbon width, respectively. Calculated electronic properties indicate that both, the SiNRs and SWSiNTs could be fully compatible with contemporary silicon-based microelectronics and could serve as natural junction and active element in field of nano-micro technologies.

The paper is divided into 4 sections. In section I.Introduction a survey of published experimental and theoretical results concerning 1D silicon nanotubes, 2D ribbons and single-layer sheet are presented. Short sketch of helical symmetry used for SWSiNT formation and band structure calculation is presented in section II.1.Helical symmetry of nanotubes in band structure calculation. The modified Hartree-Fock SCF method for periodic solids which is used for band structure calculations is shortly introduced in section II.2.The HF-SCF cyclic cluster method for band structure calculation. The results obtained at study of silicon single-layer sheet and ribbons are presented and discussed in subsection III.1.Band structure of silicene sheet and silicene ribbons. In subsection III.2 Band structures of SWSiNTs, the results obtained at study of SWSiNTs are presented and discussed. Summary of presented results are in section IV. Conclusions.

## II. Methods
## II.1. Helical symmetry of nanotubes in band structure calculation

The basic idea of accounting for the helical - screw symmetry in nanotubes as suggested in[38-40] and somewhat differently in[41], was closely followed in our implementation. Since some technical details are different and can be written in a simplified manner, we repeat them here in order to clarify the calculated band structures.

In general, any nanotube with a periodic structure can be constructed by rolling up a single sheet (ribbon) of a two-dimensional structure that is finite in one translation direction and infinite in the other one. We shall restrict ourselves to nanotubes created from two-dimensional hexagonal lattice characterized by two equivalent $|a|=|b|$ primitive translational vectors $a$ and $b$ that contain an angle of $2\pi/3$. In particular, real lattice unit vectors are $a = -aY$, $b = (\sqrt{3}/2)aX+(1/2)aY$ and corresponding reciprocal lattice vectors are $a^*=(2\pi/a)(1/\sqrt{3})X-(2\pi/a)Y$, $b^*=(2\pi/a)(2/\sqrt{3})X$ with coordinates of the high symmetry points in $k$-space $\Gamma\equiv(0,0,0)$, $K\equiv(-1/3,2/3,0)$, $M\equiv(0,1/2,0)$.

Due to our convention, the translations along the direction of $a$ will be treated as finite, whereas "infinite" number of translations is assumed along $b$. A nanotube characterized by a general helical vector ($m_a\,a + m_b\,b$) – with a tube notation $[(m_a,m_b), m_{tr}]$, is then created from the ribbon that has $m_a$ translations $(0, \ldots, m_a\text{-}1)$ along $a$ and "infinite" number $m_{tr}$ of translations along $b$. The finite value $m_b$ ($m_b \ll m_{tr}$) of helical vector number along $b$ and finite $m_a$ value characterize the first complete thread of the helix. Such a ribbon is rolled up on a cylinder with the diameter

$$d_{NT}=|m_a a+m_b b|/\pi, \qquad (1)$$

which follows from the fact that the helical vector ($m_a\,a + m_b\,b$) is rolled up perpendicular to the rotation axis and makes the circumference of the cylinder. The *irreducible computational tubular unit cell* corresponds to that in the two-dimensional structure except for the geometry relaxation due to the curvature. Exactly as in the two-dimensional structure that is infinite in both dimensions, each such unit experiences the same environment. Original translations along $a$ and $b$ are now transformed to rototranslations ($\hat{\tau}_a, \hat{\tau}_b$) characterized by the pair of operations ($z_a, \varphi_a$) and ($z_b, \varphi_b$), where $z_a, z_b$ are projections of $a$ and $b$ onto the axis of the nanotube and $\varphi_a, \varphi_b$ is the rotation angle related to this translation. Hence for any point defined in a cylindrical coordinate system ($\rho, \phi, z$) is,

$$\hat{\tau}_i \equiv (\rho, \phi+\varphi_i, z+z_i), \quad i = a, b. \qquad (2)$$

If we relate a pseudo-vectors $t_i$ to these rototranslations, in analogy with the two-dimensional planar lattice we can define reciprocal pseudo-vectors $t_i^*$ such that

$$t_i^* t_j = 2\pi\delta_{ij} \qquad (3)$$

Let the atomic orbital $\chi_{j_a,j_b}$ be a counterpart of the reference unit cell atomic orbital $\chi_{0,0}$

in the unit cell defined by $\hat{\tau}_a^{j_a}$ and $\hat{\tau}_b^{j_b}$ rototranslations. The structure created by $m_a$ rototranslations $\hat{\tau}_a$ (including 0) of the reference computational cell can be treated as an ideal cyclic cluster with periodic boundary conditions, since, indeed in the nanotube each unit has an equivalent surrounding. Consequently, from $m_a$ atomic orbitals $\chi_{j_a,0}$ ($j_a = 0$, $m_a - 1$) one can create $m_a$ symmetry orbitals:

$$\chi^{\Phi_r} = \frac{1}{\sqrt{m_a}} \sum_{j_a=0}^{m_a-1} e^{i k_{\Phi_r} \cdot R_{j_a}} \chi_{j_a,0} \quad , \tag{4}$$

where $R_{j_a} = j_a t_a$ and there are $m_a$ allowed discrete values of $k_{\Phi_r} = (r/m_a) t_a^*$ for $r = 0,..,m_a-1$. These symmetry orbitals are propagated due to the $\hat{\tau}_b$ to "infinity" ($m_{tr} = N \gg m_a, m_b$) providing Bloch orbitals:

$$\chi^{(k_{tr},k_{\Phi_r})} = \lim_{N \to \infty} \frac{1}{\sqrt{m_a \cdot N}} \sum_{j_b=-N/2}^{N/2} \sum_{j_a=0}^{m_a-1} e^{i(k_{tr} \cdot R_{j_b} + k_{\Phi_r} \cdot R_{j_a})} \chi_{j_a,0} \tag{5}$$

where $R_{j_b} = j_b t_b$ and $k$'s are any values from the first Brillouin zone for the one-dimensional system. In the practical implementation into the codes that generate integrals in a Cartesian coordinate system, *one has to take care for the appropriate rotation of the coordinate system for each basis function center to preserve the rotational symmetry*. Expressed explicitly, to rotation and rototranslation operations have to be subjected not only nuclear centers that results in set of nuclear coordinates of involved atoms on surface of tube, but to these operations all basis functions of involved atoms must be subjected as well, in order to ensure correct directional – radial and angular arrangements of p,d,f-AOs on tubular surface.

In case of nanotubes with helical vectors $[(m_a,0),m_{tr}]$, $\mathbf{k}_{\Phi r} = k_{\Phi r} t_a^*$ is related to the true rotational angle in the $m_a$-fold symmetry, whereas $\mathbf{k}_{tr} = k_{tr} t_b^*$. In this special case, the rotational symmetry can be as well accounted for through the point symmetry operations as in Ref.[41]

**II.2. The HF-SCF cyclic cluster method for band structure calculation**
The band structures have been calculated by computer code Solid2000-NT. The code is based on the Hartree-Fock SCF (HF-SCF) method for infinite periodic cyclic 3D cluster[43] with the quasi-relativistic INDO Hamiltonian[44]. Based on the results of atomic Dirac-Fock calculations[45], the INDO version used in the SOLID package is parametrized for nearly all elements of the Periodic Table. Incorporating the INDO Hamiltonian into the cyclic cluster method (with Born-Karman boundary conditions) for electronic band structure calculations has many advantages and some drawbacks as well. The method is not very convenient for strong ionic crystals but it yields good results for intermediate ionic and covalent systems. The main disadvantage is an overestimation of the total width of bands which is an inherent feature of the HF-SCF in application for periodic solids. On the other hand, it yields satisfactory results for properties related to electrons at the Fermi level (frontier-orbital properties) and for calculation of equilibrium geometries[46-48].

In practical calculations, the basic cluster of the dimension $N_a \times N_b \times N_c$, is generated by corresponding translations of the unit cell in the directions of crystallographic axes, *a* ($N_a$), *b* ($N_b$), *c* ($N_c$). In particular, the tubular band structure calculations have been performed for the basic cyclic clusters [($m_a \times m_b \times 1$), $m_{tr}$]. The basic cyclic cluster of the particular size generates a grid of ($m_a \times m_{tr}$) points in *k*-space. The HF-SCF procedure is performed for each *k*-point of the grid with the INDO Hamiltonian matrix elements that obey the boundary conditions of the cyclic cluster[43]. The Pyykko-Lohr quasi-relativistic basis set of the valence electron atomic orbitals (i.e. 3s, $3p_x$, $3p_y$, $3p_z$ -AO for Si) has been used. In the case of SWSiNT, there are 2 Si atoms and 8 AO in 2D-hP unit cell of parent silicene single-layer sheet. The basic cyclic cluster, e.g. for SWSiNT [(31,0), 91] with 5642 atoms, generates a grid of 2821 points in *k*-space and the total number of STO-type functions in the cluster is 22,568. The number of STO-type functions is unambiguously determined by the number of AOs of the valence electrons pertaining to atoms which constitute the basic cluster. Dimension of the basic cluster directly determines the number of generated *k*-points in the grid, i.e. ($m_a \times m_{tr}$). However, what is important to be stressed, is the fact that no matter of the number of the AOs in the calculated cyclic cluster, the number of the bands remains always 8 as it corresponds to parent 2D-hP unit cell of silicene sheet. That is the consequence of helical method used at tube formation since irreducible 2D-hP unit cell remains also the unit cell (irreducible) of tubular structure.

That is the crucial and substantial difference in comparison with conventionally used chiral-vector method of tube formation, e.g. as used for carbon nanotubes.[35,36] If chiral method were used for e.g. (31,0) tube construction, which is equivalent to [(31, 0), $m_{tr}$] tube within the helical treatment (diameter ≈ 38.81Å ), then chiral – translational tubular unit cell would contain 62 irreducible unit cells of 2D-hP character, i.e.124 Si atoms and band structure would be a bundle of 496 bands. In light of that, the question if tubular band structure calculated within chiral-vector treatment of tube formation represents a true tubular band structure is legitimate. Reliable answer to this question can be obtained from experimental momentum distribution curves of highly-resolved ARPES of single-wall nanotubes which are, however, unavailable so far.

The precision of the results of band structure calculation within the cyclic cluster treatment increases with increasing dimension of the basic cluster. It has been shown[43,46-48], however, that there is an effect of saturation, a bulk limit beyond which the effect of increasing dimension on e.g. total electronic energy, orbital energies, HOMO-LUMO difference..., is negligibly small. In practice, dimension of the basic cluster and parameters selection (e.g. for calculation of β integrals) is a matter of reasonable compromise between computational efficiency and compatibility of calculated electronic properties and equilibrium geometry with respect to some reference or experimental data. In the present study of silicon compounds, the scaling parameter 1.0 (standard INDO scaling of β integrals for molecular system calculations is 0.75) has been used in calculations of the one-electron off-diagonal two-center matrix elements of the Hamiltonian (β-"hopping" integrals). Using this scaling parameter, the best agreement with experimental data of bulk silicon (diamond-like cF8 structure) has been reached. Calculated band gap is 1.242 eV (exp. 1.166 eV) and equilibrium Si-Si distance is 0.240 nm (exp. 0.235 nm), allowed direct excitation energy is 0.826 eV and lowest spin-flip excitation is also 0.826 eV It should be kept in mind, however, that the basic efficiency and accuracy are restricted by the INDO method parametrization and character of the HF-SCF method itself.

### III. Results and discussion
### III.1. Band structure of silicene sheet and silicene ribbons

As mentioned in the Introduction, the single layer sheet of silicene with 2D-hP honeycomb structure can be considered as a parent structure for SWSiNT formation. The DFT-based simulations[21], yield strictly planar and low-buckled honeycomb structures with nearly the same energetic minima, but for planar type the phonon dispersion reveals hybridization of some optical and acoustic phonon branches with lowering into acoustical region and with imaginary frequency at Γ point. The low-buckled structure yields very small (zero at Γ point) but positive frequency of this branch all over Γ-K direction. Both structures are characteristic with the same optical phonon branch, with frequency ≈ 600 cm$^{-1}$ at Γ point.

The HF-SCF method used in the present study yields only slight puckering which is only several meV more stable than planar structure. Since the topology of band structure of both structural types is basically identical we use, in what follows, the planar type as a parent structure for SWSiNTs formation. Two Si atoms (Si1: 1/3, 2/3; Si2: 2/3, 1/3) constitute the unit cell with optimized lattice parameter a =b = 3.933 Å and equilibrium Si-Si distance is 2.270 Å, which is in good agreement with experimental[19] value 2.2 Å and published theoretical[20-29] DFT and TB results, i.e. 2.2 – 2.4 Å.

The band structure of planar silicene sheet calculated for basic cyclic cluster (91x91x1) is in Fig.1a.

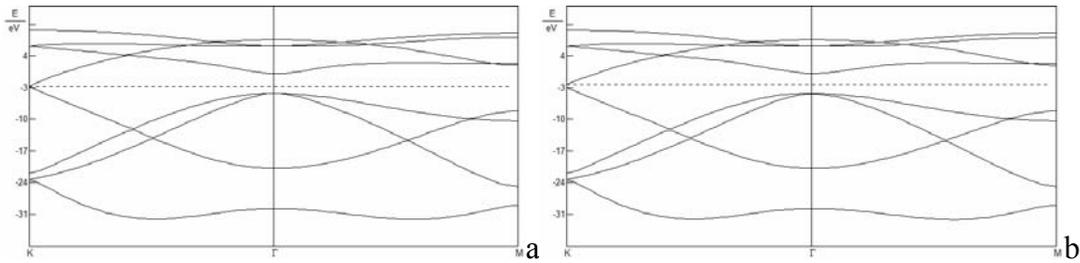

Fig. 1 Band structure of "infinite" ($N_a=N_b=91$) silicene sheet (a) and band structure with a gap at K-point of silicene nanoribbon "infinite" in *a*-direction ($N_a=91$) and finite ($N_b=5$) in *b*-direction (b). Doted line indicates Fermi level.

As it can be expected, topology of the band structure of silicene - Fig.1a, is similar to that of graphene. Dispersion of ($\pi^*$, $\pi$) bands at K-point is linear with "massless" relativistic Dirac fermions. Fermi-level velocity ($v_F=E/\hbar k$) calculated from the band structure is 2.1 10$^6$ m/s. Comparing to $v_F$=1.3 10$^6$ m/s calculated from the experimental ARPES dispersion[19], Fermi velocity calculated from the band structure is overestimated which is a consequence of exaggeration of the total band-width within the HF-SCF method.

An interesting result[19] of the experimental ARPES is the fact that the apex of the π-band dispersion is ≈ 0.3 eV below Fermi level. It indicates that studied silicene sample was rather a small-gap (≈0.6 eV) semiconductor and not a semimetal (zero-gap semiconductor) as it should correspond to ideal "infinite" single-layer sheet with relativistic Dirac fermions.

It is well known that on the nanoscale, electronic properties are very sensitive to aspects of dimensionality - quantum confinement effect.

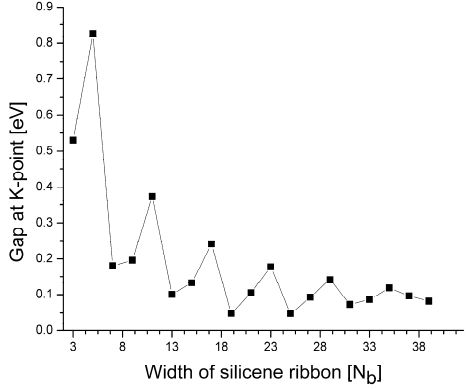

Fig.2 Dependence of calculated gap of silicene nanoribbons as a function of ribbon width $rw = N_b a \cdot \sqrt{3}/2$ for fixed ribbon length $rl = N_a a$. Parameters of the ribbons are; $N_a$= 91 for $N_b$= 3 – 39 and lattice constant $a$=3.933 Å

In Fig.2, we present dependence of calculated gap between ($\pi^*$, $\pi$) bands at K-point as a function of ribbon width ($rw = N_b a \cdot \sqrt{3}/2$), expressed over the number of $N_b$ translations of 2D-hP unit cell in **b** axis direction, going from "infinite" silicene sheet ($N_a=N_b=91$) down to a very narrow nanoribbon ($N_a=91$, $N_b=3-39$, with odd integers). As it can be seen, the overall tendency is gap decreasing with increasing ribbon width, reaching zero-gap value in the limit of infinite 2D-hP layer - ($N_a=N_b=91$). An interesting aspect is a kind of oscillatory behavior of gap dependence for neighboring triads, i.e. gap(3.($N_b$-2))< gap(3.$N_b$) < gap(3.($N_b$+2)). Similar oscillatory dependence has been reported[21] for triads 3p, 3(p+1), 3(p+2) – p integer, in armchair-type (chiral-method convention) of bare and H-saturated silicene nanoribons within DFT-based simulation. We note that our results are obtained for zig-zag type of silicene nanoribons. Calculated gap-dependence should be an explanation of the observed ARPES results. As an example, the band structure of the nanoribbon $N_a=91$, $N_b=5$; i.e. length ≈35.8 nm and width ≈1.7 nm, with opened gap 0.82 eV at K-point (excitation energy 0.278 eV) is presented in Fig.1b.

**III.2. Band structure of SWSiNTs**
Helical method of tube formation and standard crystallographic convention of hP lattice (**a,b** angle $2\pi/3$, the fractional coordinates of atoms in the unit cell; Si1: 1/3, 2/3, 0 ; Si2: 2/3, 1/3, 0) result in substantial difference of tubular band structure and in different appearance of the termination (ends) of tube comparing to conventional chiral-vector treatment (**a,b** angle $\pi/3$, the fractional coordinates of atoms in the unit cell; Si1: 1/3, 1/3, 0 ; Si2: 2/3, 2/3, 0) which is usually applied in study of carbon nanotubes. Until the chiral convention leads into armchair appearance of tubular ends for (n,n)-type structure and zig-zag edges for (n,0)-type, in case when crystallographic convention for hP lattice and helical method is used, the appearance of tube-terminations is different. For helical type $[(m,m),m_{tr}]$, which has the same radius as chiral $(m,0)$-type, ends of tube are also of zig-zag appearance but terminal atoms lay in a plane which is oblique with respect to the tube axis. However, for helical type $[(m,0),m_{tr}]$ (with the same radius as $[(m,m),m_{tr}]$-type within the helical treatment), terminal atoms lay in a plane which is perpendicular to the tube axis but ends of tube are again of zig-zag appearance. It should be remarked that to a chiral armchair tube $(n,n)$ corresponds a helical tube of the type $[(2n,n),m_{tr}]$ with zig-zag termination. This is the consequence of crystallographic hP lattice primitive vectors convention and of applied screw-symmetry operations (rototranslations) at helical method of tube formation. As an illustrative example, the SWSiNT of different helicities are presented in Fig.3a,b.

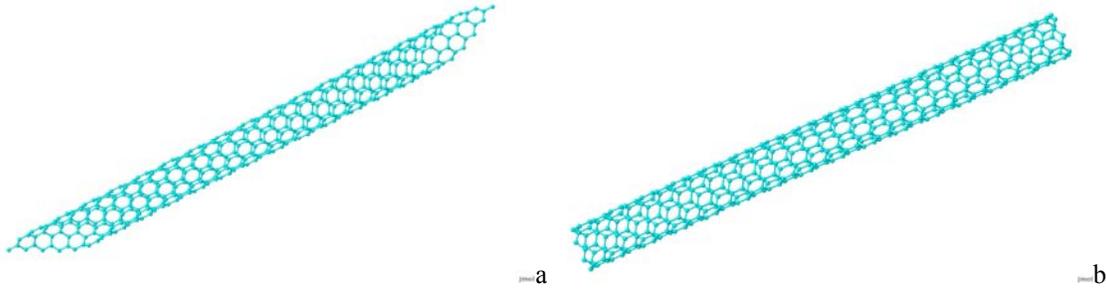

Fig.3 The appearance of the tube terminations for (a) $[(7,7), m_{tr}=31]$ and (b) $[(7,0), m_{tr}=31]$ helical types of SWSiNTs with the same diameter ≈ 8.76 Å

Substantial differences are in calculated band structures. The crucial reason is the fact that within the conventional chiral-vector treatment, tubular (translational) unit cell is not the irreducible unit cell of tubular structure. It contains N irreducible unit cells of parent 2D-hP structure. In case of silicene (graphene) it means 2N times more atoms in tubular translation unit cell comparing to parent 2D-hP unit cell. For silicene and/or graphene types $(n,0)$ and $(n,n)$ in chiral treatment notation N=2n, number of atoms is 4n and number of bands 16n. It not only increases number of bands from 8 to 8N (in valence electron basis set) but band structure itself is N-times folded in particular $k$-direction. It is clear that in those circumstances relation to band structure of parent 2D-hP is lost and can hardly be reconstructed, mainly in case of complex compounds with more than 2 atoms in 2D unit cell, e.g. boron nanotubes with 8-B atoms in 2D-hP unit cell.[27] The smallest diameter of so far synthesized SWSiNT[14-16] is about 20 Å. It roughly corresponds to (21,0)-type of SWSiNT. Within the conventional chiral treatment, it yields band structure with 672 bands. In contrast to that, due to screw-symmetry operations, the irreducible 2D-hP unit cell remains also the unit cell of tubular structure and no matter of helical vector parameters $[(m_a,m_b),m_{tr}]$, tubular and parent 2D-hP band structures within the helical treatment of tube formation are represented by the same number of bands. For valence basis set, in case of SWSiNT (or SWCNT) it is always 8 bands.

The next point which has to be mentioned is zone-folding method[35,36] frequently used for tubular band structure calculations of graphene-like structures. According to this, tubular band structure dispersion $E^{tb}(K_1,K_2)$ is derived directly from the planar 2D-hP graphene-sheet dispersion $E^{2D}(k_x,k_y)$ by simple replacement of planar 2D-hP $k$-values $(k_x,k_z)$ by tubular $k$-values $(\mu, k_{tr})$, i.e. $E^{tb} = E^{2D}(\mu, k_{tr})$. Now, however, $\mu$ is a set of discrete values related to the number N of 2D-hP unit cells in tubular unit cell, i.e. ($\mu=0, 1,…,(N-1)$) and $k_{tr}$ is continuous in the range $(0, \pm\pi/T)$ with T standing for translation modulus of tubular unit cell. For strictly planar graphene sheet with sp$^2$ hybridization, the π orbitals (p$_z$ –AO) are perpendicular to σ-orbitals (hybridized s, p$_x$, p$_y$-AOs) which lay in plane of graphene sheet and both sets of orbitals are due to symmetry reason decoupled. Within the tight-binding method and nearest-neighbor approximation it enables to derive simple analytic expression for π-band dispersion, which is then used for tubular band structure calculations by zone-folding method as usually presented in numerous publications related to electronic structure of carbon and/or graphene-like nanotubes. In planar graphene layer, the π bands are near to Fermi level and are dominant for electronic and optical properties, whilst little attention is paid to σ-bands which are more distant from the Fermi level. But, application of this simple picture, namely strict sp$^2$ hybridization and zone-folding treatment (focusing mainly on π bands), need not be the right choice in study of electronic structure of all graphene-like nanotubes. Direct calculations of tubular band structure on first-principles DFT level have revealed that calculated zone-folding band structures, mainly for small-diameter tubular structures, do not correspond to the reality. Striking differences are mainly for nanotubes of more complex compounds, e.g. for boron nanotubes with 8 –B atoms in 2D-hP lattice (Bα8 structure) where band structure for a given chirality (n,n) is metallic within zone folding treatment but of semiconductor character when band structure is calculated directly for nanotube.[27] The reason of this is the effect of sheet folding, which is not reflected in zone-folding approximation based on planar 2D-hP band structure (strict π-σ orbitals decoupling), just calculated for tubular $k$-values (μ, k$_{tr}$). In real tubular systems, mainly with strong curvature (small-diameter tubes), the strict decoupling of π-σ orbitals is not valid any more and, character of system becomes rather a kind of sp$^2$-sp$^3$ hybridization mixture.

In general, unlike of carbon with stable bulk solids of sp$^2$ character (graphite), silicone prefers structures that are rather of sp$^3$ character. Stable is diamond-like cF8 structure of bulk silicon but, silicon analogue of graphite is not known. Also recently synthesized single-layer silicene sheet[19], though of honeycomb structure, is slightly buckled.

In this respect, the effect of curvature on band structure of SWSiNTs can be expected to be even stronger than in case of carbon nanotubes. The effect of zone-folding is very pronounced for tubular band structure calculation within helical method of tube formation with irreducible unit cell, when direct correspondence between tubular band structure and parent 2D-hP band structure is preserved. In Fig. 4, band structures of small-diameter, $\phi ≈ 3.75$ Å - [(3,0),91] and $\phi ≈ 6.26$ Å - [(5,0),91] SWSiNTs are presented. As it can be expected, zone-folding band structures of (3,0)- Fig.4a and (5,0)-Fig.4c, obey simple metallicity condition [(n-m)=3μ], i.e. tube is metallic if (n-m) is multiple of 3 and semiconducting otherwise. In the presented figures, two panels (K-G-M) on the left side represent band structure of parent single-layer planar 2D-hP silicene sheet and following panels are zone-folding band structures for allowed discrete helical tubular $k_\phi$ -"vector" values $k_{\phi r}= r/m_a$ along translation path (0,0) →(0,1/2) for $r = 0$ and for [($r/m_a$,1/2)→($r/m_a$,0) | (-$r/m_a$,0) →(-$r/m_a$,1/2)], with $r ≠ 0$ arrangements. For (3,0)-SWSiNT, $r = 0,±1$, $m_a$=3 and for (5,0)-SWSiNT, $r = 0,±1,±2$, $m_a$=5. As it can be expected for zone-folding method, the band structure topology for $r$=0 is identical with that in G(≡Γ)-M for parent 2D-hP (cf. G-M↔0-M in Fig.4a and 4c). Metallicity, or rather semimetal character (formation of "Dirac cone"-like topology, direction 1-r in Fig.4a), for (3,0)-SWSiNT is due to dispersion of ($\pi^*$, $\pi$) bands for r=1, i.e. $k_{\phi r}$= - 1/3 (formally equal to the value of K-point coordinate of 2D-hP in $\boldsymbol{b}^*$ direction).

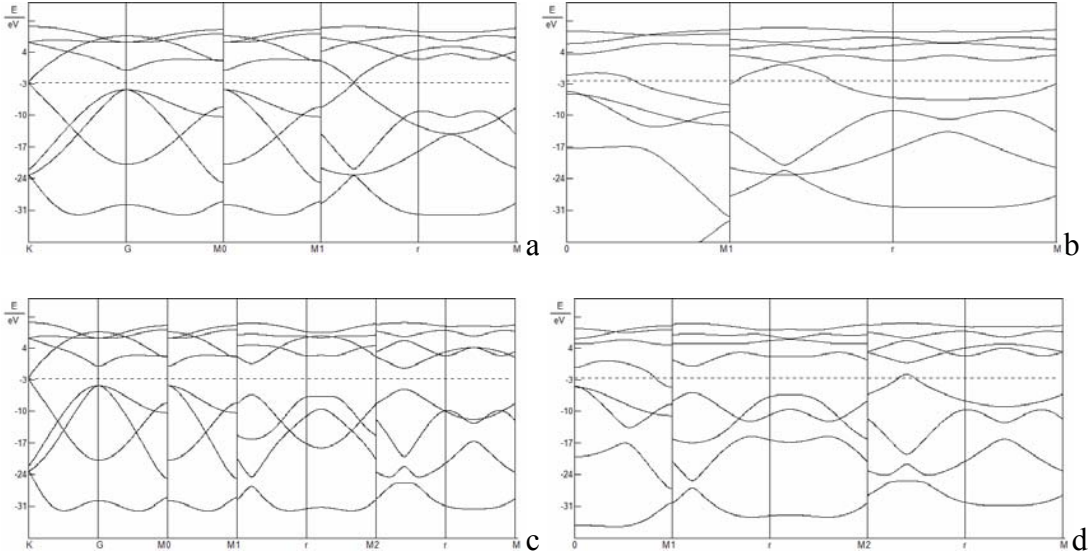

Fig.4 Band structure of (a,b) [(3,0),91], $\phi \approx 3.75$ Å and (c,d) [(5,0),91], $\phi \approx 6.26$ Å SWSiNTs. Comparison of band structures calculated by (a,c) zone-folding method and (b,d) calculated directly by helical method of tube formation with account for effect of tube curvature – for details see text.

Band structures calculated directly; i.e. when curvature effect is naturally incorporated over screw-symmetry operations, reveal dramatic changes in band structure topology for both small-diameter SWSiNTs. In panel (b) of Fig.4, the band structure of (3,0) SWSiNT with metal-like character is displayed. As it can be seen, one can hardly assignee metalicitty of this tube in 0-M direction, i.e. path (0,0)→(0,1/2), to a pure $\pi$-band. It looks rather like bended $\sigma^*$ band which from the antibonding region decreases toward Fermi level, crosses it and at M-point is sank below it. Metallicitty in direction 1-r-M, i.e path [(-1/3,1/2)→(-1/3,0) | (1/3,0) →(1/3,1/2)] in panel (b) is due to a band which is again a kind of mixture of $\pi$-$\sigma$ character, in spite of fact that band crossing above Fermi level resemble the shape of $\pi^*$-$\pi$ Dirac-like cone. Strong curvature effect persist also in (5,0) SWSiNT – panel (d) in Fig.4. Surprisingly, this tube is also metallic; in 0-M direction, i.e. path (0,0)→(0,1/2), and in 2-r-M direction, i.e. path (-2/5,1/2)→(-2/5,0) | (2/5,0) →(2/5,1/2). Whilst, like for (3,0) tube, metalicitty in 0-M direction resembles bending of $\sigma^*$ band, the metalicity in 2-r-M direction is more of $\pi$-band character.

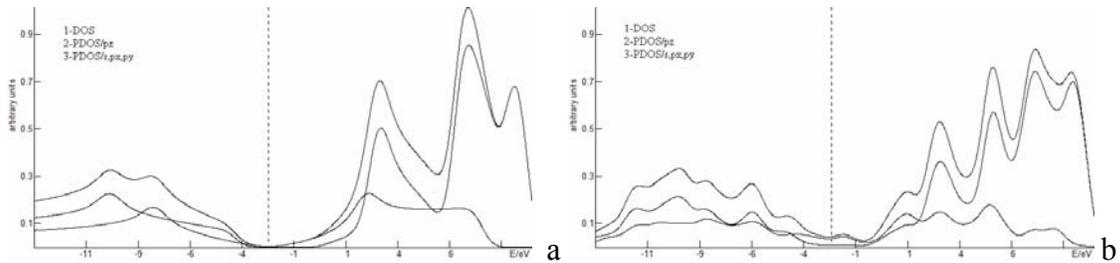

Fig.5 Calculated DOS for (a) silicene sheet and (b) [(5,0),91]-type SWSiNT with splitting on PDOS contributions from $\sigma$ and $\pi$ bands. Vertical dotted line indicates Fermi level.

Indeed, contribution to total density of states (DOS) at Fermi level from σ bands (2s,2p$_x$,2p$_y$) is not negligible; PDOS$_σ$ is ≈18% and PDOS$_π$ from p$_z$ orbitals is ≈72%. In Fig.5, density of states (DOS, PDOS$_π$ and PDOS$_σ$) for silicene sheet (a) and [(5,0),91]-SWSiNT (b) are presented. In overall, however, the DOS at Fermi level for [(5,0),91]-SWSiNT is very small, ≈0.044.

Topology of the band structure of [(5,0),91]-SWSiNT is very sensitive to electron-vibration coupling. Already at displacement of 0.039 Å/Si-atom out of equilibrium at vibration motion in stretching Si-Si mode, instability of the tubular band structure with characteristic fluctuation of band structure topology at Fermi level is induced. In particular, in this vibration displacement, degeneracy of σ-bands below Fermi level at Γ(0)-point at equilibrium geometry is lifted and maximum (an analytic critical point - ACP) of the upper σ-band is shifted above Fermi level. In vibration motion it represent periodic fluctuation of ACP across the Fermi level – cf. panels (a),(b) in Fig.6 for path 0-M.

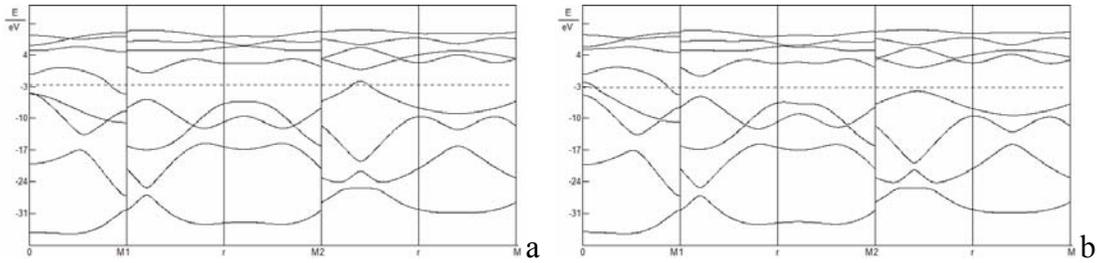

Fig.6 Effect of electron-vibration coupling on band structure topology at Fermi level for [(5,0),91]-type SWSiNT. Band structure at (a) equilibrium undergoes topology change at Fermi level for (b) displacement of 0.039 Å/Si-atom out of equilibrium at vibration motion in stretching Si-Si mode, cf. path 0-M for a/b panels.

This type of band structure fluctuation was observed for the first time in superconducting MgB$_2$[49,50] and it is related to the breakdown of the Born-Oppenheimer approximation.[51-53] The mentioned aspect of band structure instability is important for crossing of the system from adiabatic metal-like state into anti-adiabatic state which, if stabilized, is directly related to superconducting state transition.[42,51,54-56] Study of this problem, i.e. aspects of anti-adiabatic state stabilization and superconducting state transition, is out of the scope of the present paper and it will be published elsewhere.

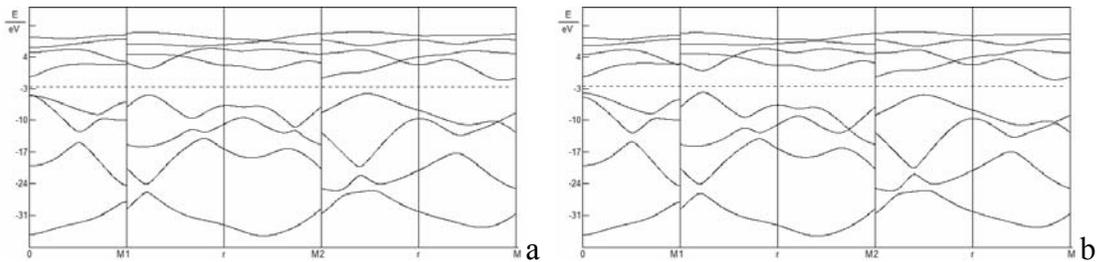

Fig.7 Effect of electron-vibration coupling on band structure topology at Fermi level for [(5,5),91]-type SWSiNT. Band structure at (a) equilibrium remains without change of topology at Fermi level (cf. path 0-M for a/b panels) for (b) vibration motion in stretching Si-Si mode, which induces topology change in [(5,0),91]-type SWSiNT, cf. Fig.6b/Fig.7b.

The SWSiNTs of the helicity [(5,3),91] - $\phi \approx 5.46$ Å, and [(5,5),91] with the same radius $\phi \approx 6.26$ Å as the [(5,0),91] type, are semiconductors with indirect gap. In Fig. 7, the band structure of [(5,5),91]-SWSiNT at equilibrium (a) and distorted geometry in Si-Si stretching mode displacement (b) is displayed. As it can be seen, the same displacement amplitude, which for (5,0)-type induces band structure instability, in spite of σ-band splitting at Γ(0)-point, leaves this tube in adiabatic state without topology change at Fermi level – maximum of σ-band does not cross Fermi level.

As an illustration of the fact that simple metallicity condition [(n-m)=3μ] related to ($\pi^*$, π) bands topology do not hold for band structures which incorporate the effect of folding, we present the band structures of the SWSiNTs with larger diameters, i.e. with smaller curvature than in case of the [(3,$m_b$),91] and [(5,$m_b$),91] tubes. In Fig. 8, band structure of the [(21,0),91]-type SWSiNT is displayed for full set [($r/21$,1/2)→($r/21$,0) | (-$r/21$,0) →(-$r/21$,1/2)] of allowed $k_\phi$, with arrangements $r = 0$, $r = \pm 1, \pm 2,... \pm 10$.

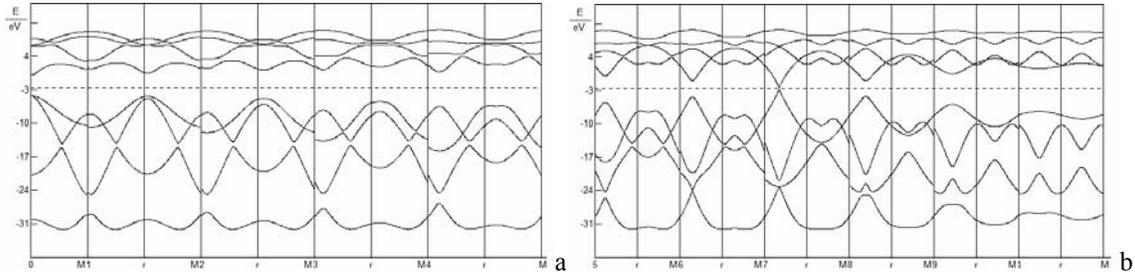

Fig.8 Band structure of [(21,0),91] SWSiNT for full set of allowed $k_\phi$, i.e. [(-$r/21$,1/2)→(-$r/21$,0) | ($r/21$,0) →($r/21$,1/2)], with arrangements (a) $r = 0$, $r = \pm 1, \pm 2, \pm 3, \pm 4$ and (b) $r = \pm 5, \pm 6,... \pm 10$.

The SWSiNTs of the type [(21,$m_b$),91; odd $m_b$], are metallic within zone-folding approximation (not displayed) for $m_b$ = (0, 3, 9, 15, 21) and semiconductors for $m_b$ = (1, 5, 7, 11, 13, 17, 19). However, due to the nanotubes curvature, the band structures of the all of SWSiNTs of the type [(21,$m_b$),91], no matter of $m_b$ value (odd) are of small-gap semiconductor characters (0.46–0.78 eV, see Table 1). Similar situation is for SWSiNTs of the type [(23,$m_b$),91]. Now, metallic within the zone-folding approximation (not displayed) are SWSiNTs for odd $m_b$ = (5, 11, 17). But, again, due to the real nanotubes curvature, the band structures of the all of SWSiNTs of the type (23,$m_b$) are of small-gap semiconductor characters (1.31–1.39 eV, see Table 1). In Fig.9, the band structure of SWSiNT of the type (23,5),91 is displayed.

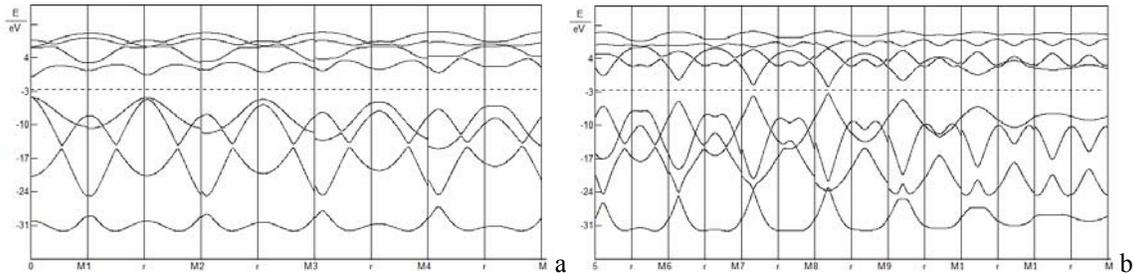

Fig.9 Band structure of [(23,5),91] SWSiNT for full set of allowed $k_\phi$, i.e. [(-$r/23$,1/2)→(-$r/23$,0) | ($r/23$,0) →($r/23$,1/2)], with arrangements (a) $r = 0$, $r = \pm 1, \pm 2, \pm 3, \pm 4$ and (b) $r = \pm 5, \pm 6,... \pm 11$.

All studied SWSiNTs with diameter $\phi > 7$ Å are semiconductors without indication of instability toward electron-vibration coupling. Dependence of the mean gap on average value of diameter (averaged over $m_b$ for fixed $m_a$ in particular helical group [$(m_a,m_b)$,91]) in diameter range ≈ 3.7 Å - 61 Å is shown in Figure 10.

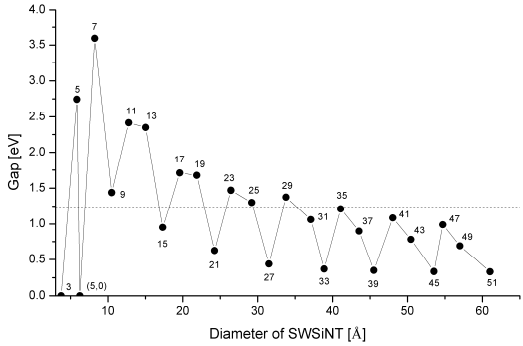

Fig.10 Calculated gap of SWSiNTs as a function of nanotube diameter. The numbers at the peaks are $m_a$-values of particular helical group which determine directly also the fold-number of particular tubular rotational axis symmetry. Horizontal dotted line indicates calculated gap (1.24 eV) of bulk silicon with cF8 diamond-like structure.

The numerals at the peaks, minima and maxima, are $m_a$-values of particular helical group. As it can be seen, the gap of SWSiNTs exhibits an oscillatory-decreasing character with increase of the tube diameter. In the oscillatory series, minima of the gap in "saw-teeth" pattern are reached for helicity numbers $m_a$ that are an integer multiple of 3 whilst, $m_a$ value itself directly determine the fold-number of tubular rotational axis symmetry for particular helical type of [$(m_a,m_b),m_{tr}$] tube. Oscillations are damped and gap decreases toward ≈ 0.33 eV for tube diameter ≈ 116 Å. It should be stressed however that gap (and diameter) within the particular $m_a$-helical group is not uniform but depends on the helical vector $(m_a,m_b)$, which by itself determines the tube circumference (diameter) and also the gap through the $m_a$-value. As an illustration, dependence of the gap for fixed $m_a$-value on running $m_b(\leq m_a)$ values for selected helical [$(m_a,m_b)$,91] groups is shown in Fig.11.

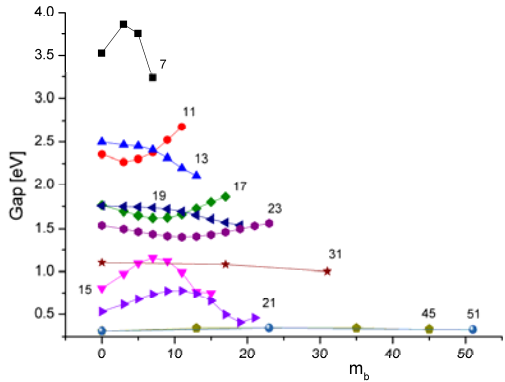

Fig.11 Dependence of the gap for fixed $m_a$-value on running $m_b(\leq m_a)$ values for selected helical [$(m_a,m_b)$,91]-types of SWSiNTs. The numbers at particular curves are $m_a$-values.

For nanotubes with larger diameters, it can be seen that for some next-neighbor helical groups $m_a$, $(m_a+2)$, depending on particular helical vectors $(m_a,m_b)$ and $(m_a+2,m_b)$, for some $m_b$ values is diameter of $m_a$-tube greater then diameter of $(m_a+2)$-type but gap in $(m_a,m_b)$ group is smaller then the gap in $(m_a+2,m_b)$ group – cf., e.g. groups $(21, m_b)$ and $(23, m_b)$ presented in Table1.

| Helicity $(n_a,m_b)$ | SWSiNT diameter [Å] | Gap [eV] | Excitation energy: allowed direct [eV] | Excitation energy: lowest spin-flip [eV] | Energy of folding/Si [eV] |
|---|---|---|---|---|---|
| (3,0) | 3.756 | metallic | 1.002 | 0.995 | 1.355 |
| (3,3) | 3.756 | metallic | 0.342 | 0.317 | 0.811 |
| (5,0) | 6.260 | metallic | 1.996 | 1.993 | 0.495 |
| (5,3) | 5.457 | 2.547 (indirect) | 2.766 | 2.763 | 0.634 |
| (5,5) | 6.260 | 2.938 (indirect) | 3.564 | 3.652 | 0.482 |
| (7,0) | 8.764 | 3.526 (indirect) | 3.384 | 3.382 | 0.220 |
| (7,3) | 7.617 | 3.857 (direct) | 3.552 | 3.550 | 0.295 |
| (7,5) | 7.819 | 3.752 (indirect) | 3.386 | 3.385 | 0.280 |
| (7,7) | 8.764 | 3.238 (direct) | 2.669 | 2.667 | 0.222 |
| (9,0) | 11.268 | 1.455 (direct) | 0.913 | 0.912 | 0.134 |
| (9,3) | 9.937 | 1.661 (direct) | 1.157 | 1.156 | 0.173 |
| (9,5) | 9.778 | 1.575 (direct) | 1.075 | 1.074 | 0.178 |
| (9,7) | 10.248 | 1.232 (direct) | 0.719 | 0.717 | 0.164 |
| (9,9) | 11.268 | 1.276 (direct) | 0.736 | 0.734 | 0.181 |
| (11,0) | 13.772 | 2.355 (direct) | 1.838 | 1.837 | 0.086 |
| (11,3) | 12.331 | 2.267 (direct) | 1.799 | 1.798 | 0.109 |
| (11,5) | 11.944 | 2.301 (direct) | 1.832 | 1.831 | 0.116 |
| (11,7) | 12.074 | 2.376 (direct) | 1.902 | 1.901 | 0.113 |
| (11,9) | 12.706 | 2.521 (direct) | 2.025 | 2.024 | 0.102 |
| (11,11) | 13.772 | 2.678 (direct) | 2.162 | 2.161 | 0.086 |
| (13,0) | 16.276 | 2.499 (direct) | 2.004 | 2.003 | 0.061 |
| (13,3) | 14.761 | 2.465 (direct) | 1.993 | 1.992 | 0.074 |
| (13,5) | 14.220 | 2.451 (direct) | 1.987 | 1.986 | 0.079 |
| (13,7) | 14.109 | 2.409 (direct) | 1.947 | 1.946 | 0.081 |
| (13,9) | 14.439 | 2.314 (direct) | 1.846 | 1.845 | 0.077 |
| (13,11) | 15.179 | 2.198 (direct) | 1.718 | 1.717 | 0.070 |
| (13,13) | 16.276 | 2.110 (direct) | 1.615 | 1.614 | 0.061 |
| (15,0) | 18.780 | 0.806 (direct) | 0.329 | 0.328 | 0.047 |
| (15,3) | 17.212 | 0.973 (direct) | 0.513 | 0.512 | 0.056 |
| (15,5) | 16.562 | 1.092 (direct) | 0.642 | 0.641 | 0.060 |
| (15,7) | 16.276 | 1.154 (direct) | 0.707 | 0.706 | 0.062 |
| (15,9) | 16.372 | 1.117 (direct) | 0.669 | 0.668 | 0.062 |
| (15,11) | 16.844 | 0.991 (direct) | 0.537 | 0.536 | 0.058 |
| (15,13) | 17.662 | 0.765 (direct) | 0.301 | 0.299 | 0.053 |
| (15,15) | 18.780 | 0.749 (direct) | 0.272 | 0.271 | 0.047 |
| (17,0) | 21.284 | 1.763 (direct) | 1.301 | 1.301 | 0.035 |
| (17,3) | 19.677 | 1.693 (direct) | 1.247 | 1.246 | 0.042 |
| (17,5) | 18.946 | 1.648 (direct) | 1.209 | 1.208 | 0.045 |
| (17,7) | 18.529 | 1.621 (direct) | 1.187 | 1.186 | 0.047 |
| (17,9) | 18.443 | 1.625 (direct) | 1.191 | 1.191 | 0.047 |
| (17,11) | 18.697 | 1.663 (direct) | 1.227 | 1.226 | 0.046 |
| (17,13) | 19.275 | 1.726 (direct) | 1.285 | 1.284 | 0.043 |
| (17,15) | 20.149 | 1.798 (direct) | 1.348 | 1.347 | 0.039 |
| (17,17) | 21.284 | 1.860 (direct) | 1.399 | 1.399 | 0.035 |
| (19,0) | 23.788 | 1.756 (direct) | 1.309 | 1.308 | 0.028 |
| (19,3) | 22.151 | 1.748 (direct) | 1.314 | 1.313 | 0.032 |
| (19,5) | 21.358 | 1.742 (direct) | 1.314 | 1.313 | 0.035 |
| (19,7) | 20.834 | 1.734 (direct) | 1.311 | 1.310 | 0.037 |

| (19,9) | 20.611 | 1.719 (direct) | 1.298 | 1.297 | 0.038 |
| --- | --- | --- | --- | --- | --- |
| (19,11) | 20.687 | 1.692 (direct) | 1.270 | 1.269 | 0.037 |
| (19,13) | 21.062 | 1.653 (direct) | 1.228 | 1.228 | 0.036 |
| (19,15) | 21.722 | 1.610 (direct) | 1.180 | 1.799 | 0.034 |
| (19,17) | 22.640 | 1.571 (direct) | 1.133 | 1.133 | 0.031 |
| (19,19) | 23.788 | 1.541 (direct) | 1.094 | 1.904 | 0.028 |
| (21,0) | 26.292 | 0.531 (direct) | 0.096 | 0.095 | 0.024 |
| (21,3) | 24.630 | 0.616 (direct) | 0.191 | 0.190 | 0.027 |
| (21,5) | 23.788 | 0.678 (direct) | 0.261 | 0.260 | 0.029 |
| (21,7) | 23.187 | 0.736 (direct) | 0.323 | 0.323 | 0.030 |
| (21,9) | 22.847 | 0.773 (direct) | 0.363 | 0.363 | 0.031 |
| (21,11) | 22.778 | 0.777 (direct) | 0.368 | 0.365 | 0.031 |
| (21,13) | 22.984 | 0.741 (direct) | 0.330 | 0.330 | 0.031 |
| (21,15) | 24.457 | 0.657 (direct) | 0.243 | 0.242 | 0.029 |
| (21,17) | 24.181 | 0.497 (direct) | 0.077 | 0.077 | 0.028 |
| (21,19) | 25.134 | 0.406 (direct) | 0.020 | 0.021 | 0.026 |
| (21,21) | 26.292 | 0.459 (direct) | 0.026 | 0.025 | 0.023 |
| (23,0) | 28.796 | 1.538 (direct) | 1.115 | 1.114 | 0.019 |
| (23,3) | 27.114 | 1.497 (direct) | 1.084 | 1.083 | 0.022 |
| (23,5) | 26.233 | 1.466 (direct) | 1.058 | 1.058 | 0.023 |
| (23,7) | 25.567 | 1.437 (direct) | 1.034 | 1.033 | 0.024 |
| (23,9) | 25.134 | 1.415(direct) | 1.014 | 1.014 | 0.025 |
| (23,11) | 24.946 | 1,405 (direct) | 1.006 | 1.006 | 0.025 |
| (23,13) | 25.009 | 1.411 (direct) | 1.011 | 1.011 | 0.026 |
| (23,15) | 25.321 | 1.431 (direct) | 1.030 | 1.029 | 0.025 |
| (23,17) | 25.872 | 1.462 (direct) | 1.058 | 1.057 | 0.024 |
| (23,19) | 26.648 | 1.498 (direct) | 1.089 | 1.088 | 0.022 |
| (23,21) | 27.629 | 1.533 (direct) | 1.118 | 1.118 | 0.021 |
| (23,23) | 28.796 | 1.563 (direct) | 1.143 | 1.141 | 0.019 |
| (31,0) | 38.812 | 1.101 (direct) | 0.718 | 0.718 | 0.011 |
| (31,17) | 33.665 | 1.079 (direct) | 0.714 | 0.714 | 0.014 |
| (31,31) | 38.812 | 1.002 (direct) | 0.622 | 0.622 | 0.011 |
| (45,0) | 56.341 | 0.314 (direct) | 0.001 | 0.001 | 0.005 |
| (45,13) | 50.221 | 0.344 (direct) | 0.020 | 0.020 | 0.006 |
| (45,35) | 51.241 | 0.344 (direct) | 0.020 | 0.020 | 0.006 |
| (45,45) | 56.341 | 0.332 (direct) | 0.003 | 0.003 | 0.005 |
| (51,0) | 63.853 | 0.314 (direct) | 0.038 | 0.038 | 0.004 |
| (51,23) | 55.387 | 0.347 (direct) | 0.040 | 0.040 | 0.005 |
| (51,51) | 63.853 | 0.329 (direct) | 0.018 | 0.018 | 0.003 |

Table 1 Calculated basic physical parameters of selected $[(m_a, m_b), 91]$ – types of SWSiNTs.

The energy of folding per Si atom presented in Table 1, is calculated as the difference between the total electronic energy of SWSiNT $[(m_a, m_b), 91]$ and total electronic energy of 2D-hP silicene single layer sheet (91x91x1), i.e., $E_{fd} = (E_{NT} − E_{sh})/2$. As it can be expected, energy expense for tube formation decreases with increasing diameter of nanotube. As far as the cohesive energy is concerned, with respect to the INDO parameterization, only relative values are relevant. Calculated relative cohesive energy per atom (at 0 K) of infinite 2D-hP silicene sheet with respect to cohesive energy of most stable bulk silicon form with cF8 structure (diamond-like) can be calculated as; $\Delta E_{r.coh} = \left(- E_{total}^{cF8}/n_{Si} + E_{total}^{2D-hP}/m_{Si}\right)$. Since silicon cF8 and 2D-hP (also SWSiNT) have the same number of Si atoms in respective unit cells then, $n_{Si} = m_{Si} = 2$. Calculated relative cohesive energy of 2D-hP silicene sheet per Si atom is $\Delta E_{r.coh} = 1.9$ eV/Si. It means that cF8 bulk silicon is more stable by1.9 eV/Si than 2D-hP silicene sheet.

Relative cohesive energies of SWSiNTs can be calculated straightforwardly by adding 1.9 eV to the corresponding energy of folding. The value $\Delta E_{r.coh} = 1.9$ eV/Si calculated within the INDO parametrization seems to be overestimated, however. Nonetheless, since single layer 2D-hP silicene sheet has been synthesized[19], i.e. this structure is thermodynamically (meta-) stable, then it can be assumed (and calculated data support it) that SWSiNTs formation should be kinetically driven process and synthesis should be a matter of tuning proper experimental conditions.

The excitation energies presented in Table 1, are calculated within the single-Slater determinant approximation optimized for the ground electronic state $E_{0,R_0}(\Psi_0)$. The direct allowed excitation energy, $E_{ex}^1 = E_{R_0}^1(\Psi_{0\to1}^1) - E_{0,R_0}(\Psi_0)$, is vertical excitation from the ground state HOMO orbital to nearest LUMO on global energy scale (allowed due to symmetry) with final singlet state configuration. The lowest spin-flip excitation energy, $E_{ex}^3 = E_{R_0}^3(\Psi_{0\to1}^3) - E_{0,R_0}(\Psi_0)$, is calculated as a lowest excitation which should be possible at some circumstances (e.g. due to electron-vibration coupling) from the ground state global HOMO orbital to nearest LUMO in a *k*-point of some path of first Brillouin zone with final triplet state configuration. It should be mentioned that within the Hartree-Fock method, calculated *k*-dependent gap energy, i.e. $gap_{R_0} = \varepsilon_{LUMO} - \varepsilon_{LUMO}$, is different from excitation energy. It is due to the fact, that total electronic energy is not the simple summation over lowest doubly occupied orbitals. The difference is due to the presence of two-electron terms (coulomb $J_{ij}$ and exchange $K_{i,j}$ integrals) in orbital energy terms $\varepsilon_i = h_{ii} + \sum_i (2J_{ij} - K_{ij})$ and consequently, have to be subtracted in total energy calculation, i.e. $E = 2\sum_i \varepsilon_i - \sum_{i,j}(2J_{i,j} - K_{i,j})$.

## IV. Conclusions

We have performed comprehensive theoretical study of silicon nanoribbons and single-wall silicon nanotubes of the length ≈ 358 Å and diameter range ≈ 3.7 Å – 116 Å with parent honeycomb structure that is compatible with structural pattering of recently synthesized silicon single-layer sheet with Si-Si distance ≈ 2.2 Å. For tubular structure construction and band structure calculations, helical treatment based on screw-symmetry operations has been used. Within this treatment, the irreducible unit cell of parent 2D-hP structure (2 Si-atoms/unit cell) remains irreducible unit cell also for tubular structure. Consequently, calculated band structure of a single-wall silicon nanotube, irrespective of diameter and helicity/chirality, is in the valence electron basis set represented always by 8 bands for particular-allowed "rotational wave number" $k_{\Phi r}$. In this way, direct correspondence with band structure of the parent 2-D hP silicon sheet is preserved. In contrast, if standard chiral-vector treatment is used, the translation tubular unit cell consist of N (N=2n, for (n,0) or (n,n) chiral vectors) irreducible unit cells of parent 2D-hP structure and calculated tubular band structure is represented by an obscure bundle of bands, which is a well known picture from usually published band structures of graphene-like nanotubes. More over helical treatment covers directly, through screw symmetry operations, curvature effect of tube formation on resulting band structure, the effect which is completely neglected within the zone-folding method. That is the reason why for the same chiral vector (n,m)-type, the basic character of band structure calculated directly for translation tubular unit cell is often different as that calculated by zone-folding method.

For study of electronic structure properties of silicon nanoribons (SiNR) and single-wall silicon nanotubes (SWSiNT), the Hartree-Fock self-consistent field method modified for periodic solids with INDO Hamiltonian and parametrization that reproduces experimental geometry and gap of bulk silicon in diamond-like cF8 structure, has been used.

The results obtained for SWSiNTs of the length of ≈ 358 Å in diameter range ≈ 3.7 Å – 116 Å of different helicity-types have shown that only small-diameter SWSiNTs up to $\phi <$ 6.3 Å are metallic due to the effect of curvature which induces coupling of σ and π orbitals. Nonetheless, these tubes - in particular [(5,0),91]-SWSiNT exhibits electronic structure instability with respect to electron-vibration coupling which is manifested by band structure fluctuation characteristic for transition into antiadiabatic state. This type of fluctuation is closely related to possibility of transition into superconducting state if antiadiabatic state is stabilized. Study of this problem is out of the scope of the present paper, however.

From the calculated band structures follow that irrespective of helicity, the SWSiNTs of larger diameter are all small-gap semiconductors with direct gap between the Dirac-like cones of ($\pi^*$, π) bands. Gap of SWSiNTs exhibits, however, an oscillatory-decreasing character with increase of the tube diameter. In the oscillatory series, minima of the gap in "saw-teeth" pattern are reached for helicity numbers $m_a$ that are an integer multiple of 3, whilst $m_a$ value itself directly determine the fold-number of particular tubular rotational axis symmetry. Oscillations are damped and gap decreases toward ≈ 0.33 eV for tube diameter ≈ 116 Å. Irrespective of the width, the SiNRs are all small-gap semiconductors, characteristic by oscillatory decreasing gap with increasing ribbon widths. From the results follows that gap of SWSiNTs and SiNRs is tuneable through modulation of tube diameter or ribbon width, respectively.

Calculated basic physical parameters, e.g. gap, excitation energy, energy of folding, relative cohesive energy, indicate that both, the SiNRs and SWSiNTs could be fully compatible with contemporary silicon-based microelectronics and could serve as natural junction and active elements in nano/micro technologies, including optoelectronics.

## Acknowledgements

This work was supported by a research grant APVV-0201-11 and partially by grant VEGA No. 1/0005/11 of the Ministry of Education of the Slovak Republic. The authors acknowledge Uniqstech a.s. for permission to use the source code of Solid2000 for this research.